\def \be {\begin{equation}}
\def \ee {\end{equation}}
\def \bea {\begin{eqnarray}}
\def \eea {\end{eqnarray}}

\def \rr {\raise.35ex\hbox{\small $\prime$}\kern-.17em{\mbox{\large $\imath$}}}
\def \del {\partial}
\def \dels {\partial\kern-.5em / \kern.5em}
\def \As {{A\kern-.5em / \kern.5em}}
\def \Ds {D\kern-.7em / \kern.5em}

\def \a {\alpha}

\def \d {\delta}
\def \eps {\epsilon}

\documentclass[10pt]{article}

\usepackage{amsfonts}

\begin{document}
\begin{titlepage}

\begin{center}
\hfill hep-th/0308103\\
\vskip .5in

\textbf{\large
Regularization of Newton Constant,
Trans-Planckian Dispersion Relation,
and Symmetry of Particle Spectrum
}

\vskip .5in
{\large Pei-Ming Ho $^a$ $^b$}
\footnote{
On leave from Department of Physics,
National Taiwan University, Taiwan, R.O.C.
}
\vskip 15pt

{\small $^a$ Department of Physics,
Harvard University,
Cambridge, MA 02138, U.S.A.}\\
{\small $^b$ Department of Physics,
National Taiwan University,
Taipei, Taiwan, R.O.C.}\\

\vskip .2in
\sffamily{
pmho@fas.harvard.edu}

\vspace{60pt}
\end{center}
\begin{abstract}

We consider the possibility that the UV completeness
of a fundamental theory is achieved by a modification
of propagators at large momenta.
We assume that general covariance is preserved at all energies,
and focus on the coupling of a scalar field
to the background geometry as an example.
Naively, one expects that the gravitational interaction,
like Yukawa interactions,
will be regularized by a propagator which decays to zero
sufficiently fast above some cutoff scale,
but we show that in order to avoid the ultra-violet divergence,
the propagator should approach to a nonzero constant.
This incompatibility between the regularizations of
gravitational and Yukawa interactions
suggests that a symmetry of the particle spectrum is needed
for a UV complete fundamental theory.

\end{abstract}
\end{titlepage}
\setcounter{footnote}{0}

\section{Introduction}

It is generally believed that a fundamental theory
including quantum gravity should be free of divergences.
In this paper we want to study the implications of
the UV finiteness of the fundamental theory.
What are the properties the theory should possess
in order to be UV finite?
Finding answers to such questions may help us
to construct the fundamental theory;
and even after the fundamental theory is found,
this kind of questions may help us understand it better.

We will assume that the UV completeness
of the fundamental theory holds perturbatively,
i.e., the theory is still UV finite after
turning off all interactions among the matter fields.
This is obviously a technical necessity.
If the UV completeness is a nonperturbative feature
of the fundamental theory,
we will not be able to derive any implication from it
until knowing the exact nonperturbative definiton of the theory.

As a first step toward understanding what is needed
for the quantum gravity to be UV finite,
we consider the gravitational coupling of a free scalar field
to the background geometry.
With the standard minimal coupling to a curved spacetime,
the quantum loop diagrams diverges,
and we would like to know what kind of modifications
to the standard theory will regularize it.

More specifically, in a curved background,
the path integral of a matter field $\phi$ generates
a Hilbert-Einstein term in its effective action.
This leads to a quantum correction to the Newton constant
\be
G^{-1} = G_0^{-1}+\delta G^{-1},
\ee
where $G_0$ is the bare coupling constant,
and $\delta G^{-1}$ is the coefficient of
the Hilbert-Einstein term in the effective action.
In the theory of induced gravity \cite{Sakharov},
$G_0^{-1} = 0$,
and the full Newton constant arises as quantum corrections.
With the standard coupling of matter fields to gravity,
$\delta G^{-1}$ diverges and needs to be regularized.

In a UV-complete theory,
both the bare coupling $G_0^{-1}$ and
the quantum correction should be finite.
In the following we will derive constraints
on the unknown fundamental theory
by requiring $\delta G^{-1}$ to be finite.

A matter field's quantum contribution to Newton constant
is closely related to that to the black hole entropy \cite{SU}.
Previous works \cite{Unruh,CJ,CCL} proposed to suitably modify
dispersion relations at high energies
to regularize the black hole entropy.
\footnote{
An exception is \cite{HB} in which Pauli-Villars
regularization was used to calculate Hawking radiation.}
Hypothetical relations between Hamiltonian and momentum
can make Newton constant finite,
but its obvious disadvantage is that general covariance is broken.
We would like to preserve the general covariance
as an exact symmetry
although it is logically possible that
it is only an approximate symmetry at low energies.
Another technical disadvantage of
modifying dispersion relations is that,
as a result of the loss of general covariance,
one needs to propose dispersion relations
separately for each class of backgrounds,
and calculations may need to be carried out independently for each case.

Instead of modifying the dispersion relation,
we consider modification to the Lagrangian by higher derivatives.
At the zeroth order approximation with respect to all other interactions,
the Lagrangian of a scalar field is quadratic,
and is characterized by the correpsonding propagator.
The finiteness of the Newton constant
imposes constraints on the propagator,
assuming that the quantum contribution of each field is finite.
Remarkably, our calculation shows that the propagator
has to approach to a nonvanishing constant
in the limit of large momentum $k^2 \rightarrow \infty$.

This is in contrast with the situation for Yukawa type interactions
in which the propagator needs to approach
to zero fast enough as $k^2 \rightarrow \infty$,
so that Feynman loop diagrams converge.
Because of this difference between gravitational and Yukawa interactions,
one can not make both Newton constant and Yukawa couplings finite
by propagator modification.

Hence we are forced to reexamine our assumptions.
A possibility is that some symmetry among all matter fields,
such as supersymmetry,
ensures that the infinities coming from different fields cancel each other.
A more complete set of alternatives will be examined
in Sec. \ref{Remarks}.

\section{One Loop Calculation}

%

In the zeroth order approximation with respect to all couplings
except the gravitational coupling to the background geometry,
the action of the fundamental theory is a sum of free field actions
for each particle field.
For a scalar field, the action is of the form
\be \label{S}
S = \frac{1}{2} \int d^d x \phi{\cal D}\phi,
\ee
where ${\cal D}$ is a pseudodifferential operator.
The effective action $W$ is defined by the path integral
\be \label{Z}
e^{-W} = \int D\phi e^{\frac{1}{2}\int d^d x \phi{\cal D}\phi}.
\ee
Note that $W$ is defined up to an (infinite) constant in the above.
Apart from this constant,
terms involving physical interactions should be finite
in a UV complete theory.

%
%



On a slightly curved space,
one can express the effective action as a power expansion
of the spacetime curvature
\be \label{W0}
W = \delta\Lambda_0 \int dV + \delta \lambda_0 \int d\sigma
+\delta G^{-1} (\int dV R - 2\int d\sigma K) +\cdots,
\ee
where $d\sigma$ is the volume element on the boundary
of the 4 dimensional spacetime,
and $K$ is the curvature on the boundary.

Due to general covariance,
the effective action $W$ is always of the form (\ref{W0})
in the weak curvature expansion,
with possibly different coefficients.

The first two terms in (\ref{W0}) correspond to
the quantum contribution of $\phi$ to
the cosmological constant in the bulk and on the boundary.
The rest are the lowest order generally invariant terms
in the weak curvature expansion,
and they contribute to Newton constant.
The standard choice of ${\cal D}$ in (\ref{S})
is the Laplace operator for a massless scalar field,
for which all the coefficients in $W$ are UV divergent.

The boundary term $\int K$ in (\ref{W0})
has to be present for the following reason.
In canonical formulation,
the Lagrangian should be a function of
the dynamical variables and their first derivatives.
Since the curvature $R$ involves second derivatives of the metric,
we should apply integration by parts to the bulk term
so that it only has first derivatives.
The boundary term in (\ref{W0}) then cancels
the boundary term generated by the integration by parts,
so that the full action is of the standard form.
For this reason the coefficient of the boundary term
is determined by that of the bulk term.
The fact that the ratio of the coefficients is fixed
has very significant physical implications.
It provides a low energy argument \cite{LW} for
the compatibility between renormalization of Newton constant
and that of the Bekenstein Hawking entropy \cite{SU}.

Let us digress a little to explain this point
as it will also shed light on possible applications
of the results of this paper to the black hole entropy.
As we just saw,
quantum fluctuations of a matter field contribute
to the Newton constant.
At the same time it contributes to the black hole entropy.
The Bekenstein-Hawking relation
\be \label{BH}
S = \frac{A}{4G}
\ee
is therefore corrected on both sides.
In \cite{SU} it was suggested that the renormalized entropy
and the renormalized Newton constant always cooperate to
keep (\ref{BH}) intact in the low energy limit.
A low energy argument for this claim was given in \cite{LW} as follows.
We saw above that the low energy effective action (\ref{W0})
determines the quantum correction to the Newton constant.
In fact $W$ also determines the entropy,
at least according to one of its definitions.
For a black hole background,
as a vacuum solution to Einstein gravity,
the Hilbert-Einstein term in (\ref{W0}) vanishes,
and $W$ is thus determined by the boundary term.
Hence the entropy is proportional to the coefficient
of the boundary term in $W$,
while Newton constant to the bulk term coefficient.
The fixed ratio of the coefficients is thus translated to
the fixed ratio of the renormalization of $S$ and $G^{-1}$.
This connection between the quantum correction of
Newton cosntant and that of black hole entropy
is part of the motivation of this work.


Now we calculate the quantum correction to Newton constant
due to a scalar field with the action (\ref{S}).
In the fundamental theory,
the operator ${\cal D}$ should produce finite coefficients in $W$.
The most general quadratic generally covariant
Lagrangian for $\phi$ allowing higher derivatives is
\be
{\cal L} = \phi f(-\nabla^2)\phi.
\ee
Although higher derivatives are notorious for
introducing various problems to the canonical formulation,
from the viewpoint of effective theories,
the proper way to deal with higher derivatives is
to treat it perturbatively \cite{YF,CHY1,CHY2},
and various problems may be resolved.


The gravitational coupling of the scalar field
to the background geometry is hidden in the inverse propagator
\be \label{fF}
f(-\nabla^2) = f(-\del^2) + {\cal F}+\cdots,
\ee
where ${\cal F}$ is the first order term
in the weak curvature expansion.
At this point we can already see the origin of
the drastic difference between gravitational coupling
and Yukawa coupling.
While Yukawa coupling is independent of the propagator,
the gravitational coupling is determined by the propagator
because of the requirement of general covariance.
A propagator that decays to zero sufficiently fast over some energy scale
effectively cuts off the integration over momentum for a Feynman diagram.
This helps the Yukawa coupling to be UV finite.
However, if the propagator is modified to have steeper slopes,
the gravitational coupling ${\cal F}$ is enhanced,
and it is not clear whether the Feynman integral will be
more divergent or less divergent.
We need to calculate the loop diagrams explicitly to decide.


An advantage of our approach is that
since the calculation is generally covariant,
we do not have to do it for every background.
We can choose any background with nontrivial curvature.
For simplicity, we choose the background to be a (large) sphere $S^d$.
In normal coordinates, the metric of a sphere is
\be
g_{ij} = (1 + \a x^2)\d_{ij} + \cdots
\ee
in the neighborhood of the north pole.
The parameter $\a$ is proportional to the constant curvature on the sphere.
One can easily check that
\be
\nabla^2 = \frac{1}{\sqrt{g}}\del_i\sqrt{g}g^{ij}\del_j
         = \del^2 + \Delta,
\ee
where
\be
\Delta = \a\left( (d-2)x\cdot\del - x^2 \del^2 \right) + \cdots.
\ee

Using the identity
\be \label{comm}
[ (\del^2)^n, \Delta ] = -4n\a\left( n + x\cdot\del \right)(\del^2)^n,
\ee
one can show that for a generic $f(-\nabla^2)$,
${\cal F}$ defined by (\ref{fF}) is of the form
\be
{\cal F} = {\cal F}_0(-\del^2) +
(x\cdot\del) {\cal F}_1(-\del^2) + x^2 {\cal F}_2(-\del^2).
\ee
We will only need the expressions of ${\cal F}_0$ and ${\cal F}_2$ below.
For any smooth function $f$,
\bea
{\cal F}_0(y) &=& \a\left(2y^2\frac{d^3}{dy^3}+3y\frac{d^2}{dy^2}\right)f(y)
=2\a y^{-1/2}\frac{d}{dy}\left(y^{3/2}\frac{d^2}{dy^2}f(y)\right), \label{f0} \\
{\cal F}_2(y) &=& \a y \frac{d}{dy}f(y), \label{f2}
\eea
where we have denoted $k^2$ by $y$.
The ${\cal F}_1$ term does not contribute because
it is odd under $k\rightarrow -k$
and we will integrate over $k$ for the one loop diagram.

It is straightforward to find
the one-loop effective action
\be
W=\frac{1}{2}\int d^d x \int\frac{d^d k}{(2\pi)^d}
\frac{{\cal F}_0(k^2)+x^2{\cal F}_2(k^2)}{f(k^2)}
+\mbox{constant}.
\ee
The first term is the Hilbert Einstein term
as it is proportional to $\alpha$,
which is proportional to the constant curvature.
The contribution of quantum fluctuations of $\phi$ to
Newton constant is thus proportional to
\be
\int d^d x (w+x^2 v),
\ee
where
\be \label{w}
w = \int dy y^{\frac{d-2}{2}} \frac{{\cal F}_0(y)}{f(y)}, \quad
v = \int dy y^{\frac{d-2}{2}} \frac{{\cal F}_2(y)}{f(y)}.
\ee
For this quantum correction to be finite,
we need both $w$ and $v$ to be finite.

We check the finiteness of $w$ and $v$ by
dividing the integral over $y=k^2$
\be
w = w_0 + w_1 + w_{\infty}, \quad
v = v_0 + v_1 + v_{\infty},
\ee
to the region of small $y$, the region of finite $y$
and the region of infinite $y$.
The middle region usually contributes a finite number to
the Newton constant for a well defined $f(y)$.
We need to examine the small $y$ region and large $y$ region
more carefully.

For the integral over $y$ around $y=0$,
we can approximate $f(y)$ by the canonical expression
\be
f(y) \simeq y,
\ee
and find
\be
w_0 = 0, 
\quad
v_0 \propto \int_0 dy y^{\frac{d-2}{2}},
\ee
which is finite if $d\geq 2$.

For the integral over $y$ to $y\rightarrow\infty$,
we need to examine various asymptotic bahaviors of $f(y)$.
Assuming that $f(y) \propto y^n$ for large $y$,
we find
\be
w_{\infty} \propto \int^{\infty} dy y^{\frac{d-4}{2}}, \quad
v_{\infty} \propto \int^{\infty} dy y^{\frac{d-2}{2}}.
\label{vinfty}
\ee
The expressions above are not valid if $n = 0, 1/2$ or $1$,
in which case $w_{\infty}$ vanishes according to (\ref{f0}).
Similarly $v$ vanishes if $n = 0$ according to (\ref{f2}).
Therefore the only case with finite quantum correction to
Newton constant is $n=0$.
That is, we need
\be
f(k^2) \rightarrow \mbox{constant} \quad \mbox{as} \quad k^2 \rightarrow \infty.
\ee
In addition, this constant has to be nonzero for $w$ and $v$ to be finite.

One can try a more general ansatz for $f(y)$ such as
\be \label{exp}
f(y) = y^n e^{\beta y},
\ee
but we will only arrive at the same conclusion
that we need $f$ to approach to a nonzero constant.
Although $0$ is also a constant and
${\cal F}\rightarrow 0$ if $f\rightarrow 0$,
the factor $y^{(d-2)/2}$ in $v_{\infty}$ (\ref{vinfty})
keeps us from finding examples for which $v_{\infty}$
is finite while $f\rightarrow 0$.

Examples of inverse propagators which give
finite quantum contribution to Newton constant include
\bea
f(-\nabla^2) &=& \frac{M^2 (-\nabla^2 + m^2)}{-\nabla^2+M^2},
\label{fraction} \\
f(-\nabla^2) &=& M^2(1-e^{\nabla^2/M^2}) + m^2, \\
f(-\nabla^2) &=& -\nabla^2 e^{\nabla^2/M^2} + m^2,
\eea
where $m$ is a constant determining the low energy effective mass of $\phi$,
and $M$ is the fundamental energy at which
the wave equation is significantly modified.
These examples satisfy the following desirable properties
\begin{enumerate}
\item The wave eqution is covariant.
\item The low energy wave equation is canonical, i.e.,
$f(k^2) \rightarrow (k^2+m_0^2)$ for small $k$.
\item The contribution of quantum fluctuation to Newton constant is finite.
\end{enumerate}

What we see here is quite surprising at first sight.
Naively, people expect that the quantum loop calculation
is UV finite if we impose a cutoff in the propagator,
that is, if the propagator goes to zero quickly
above some scale $k^2 = \Lambda^2$.
Our calculation shows however that the UV finiteness
of Newton constant requires instead that
the propagator $f^{-1}(k^2)$ approaches to a nonvanishing constant.

As we commented above,
this is due to the fact that gravitational coupling
is determined by the propagator.
When we turn on a curvature in the background,
the new interaction term $\phi{\cal F}\phi$ appearing in the Lagrangian
is larger when $f(k^2)$ varies faster with $k^2$,
because ${\cal F}$ is roughly proportional to the derivative of $f$.
The exponential decay factor in (\ref{exp}), for example,
will make both the denominator $f$ and the numerator ${\cal F}$
to increase exponentially with $y$ in the expressions of $w$ and $v$.
It is therefore not helpful to have the propagator $f^{-1}$
to vanish quickly above some scale.

Without any calculation,
one can easily see that if the propagator is a constant,
the field is decoupled from gravity.
Our calculation shows that this extreme decoupling is needed
at high energies for the UV convergence of gravitational interaction.
Once this conclusion is made,
we expect that all higher order terms in the curvature expansion
of the effective action will also be finite.

As a side remark,
since gauge field interactions are also encoded
in the kinetic term of a scalar field,
a propagator that approaches to a nonzero constant
also defines a UV finite gauge field interaction.

\section{Remarks} \label{Remarks}

%

The main result of the previous section is that
the propagator has to approach to a nonzero constant
for the Newton constant to be UV finite.
By itself the result can have many applications
to high energy physics and cosmology.
Here we discuss its implications on the logical possibilities
for a UV complete theory of quantum gravity.

\begin{enumerate}
\item \label{constant}
$f\rightarrow$ constant:

In this case the Newton constant is finite for any field content
in the free field limit of the theory.

\begin{enumerate}
\item \label{noY}
If there is Yukawa coupling in the theory,
one needs a mechanism,
possibly a large symmetry among all the fields,
to make loop diagrams involving Yukawa interactions finite.
\item
There is no Yukawa coupling in the theory.
The Yukawa coupling in the standard model arises
from other types of interactions in the high energies.
\end{enumerate}

\item \label{2}
$f\rightarrow \infty$:

In this case we need a symmetry,
such as supersymmetry, among the field content
of the fundamental theory to cancel each field's
contribution to the Newton constant.
It will be interesting to calculate higher order terms
in the effective action $W$.
A tempting scenerio is that infinitely many constraints
are arrived and the spectrum is uniquely fixed.

\item \label{3}
Neither of the above:

A symmetry or symmetries among all the fields of the theory is
responsible for making both gravitational interaction and
other interactions UV finite.

\end{enumerate}

Except the possibility (\ref{noY}),
all possibilities above imply
the necessity of a symmetry among the particles of the theory.
An inventive reader might also point out other possibilities.
Another simple possibility which admits a finite Newton constant
without requiring the propagator to approach to a constant
is to add nonminimal couplings to $R$ for some scalar fields.
However, a perfect cancellation of the divergences
from all species by the nonminimal coupling constants for
part of the particle spectrum is unnatural without a symmetry.

Furthermore, unless the propagator goes to a nonzero constant,
the UV problem for gravitational coupling
becomes more serious when we consider
higher order terms in the curvature expansion of the effective action.
As we go to higher and higher order terms,
we will need to fine tune more and more couplings.
This is apparently highly unnatural.

On the contrary, if the propagator approaches to a constant,
we can easily argue that all higher order terms will also be finite
because in the high energy limit
the free field Lagrangian is simply decoupled
from the background geometry.

The possibility (\ref{constant}) seems to provide
the simplest way to construct a UV finite theory of quantum gravity.
A toy model can be built on two scalar fields.
Gravity is induced by integrating out one of the scalar field.
Then we obtain a UV finite quantum gravity
coupled to a single scalar field.
Such toy models may be useful in the future
to understand conceptual puzzles about quantum gravity.

Another remarkable property of
the possibility \ref{constant} is that
the large $k$ behavior of the propagator is almost
completely determined.
This kind of information for trans-Planckian physics
is very precious and can be applied to cosmological models
for the very early universe.
Possible effects of higher derivatives in cosmology
have been discussed in \cite{GHR,BH,HL,TMB}.

For the possibility (\ref{2}),
the finiteness of the Newton constant leads
to a strong constraint on this symmetry we need.
For example,
the contribution of minimally coupled fields of spin $0$, $1/2$ and $1$
to the Newton constant have been calculated \cite{Kabat}
\bea
\delta G^{-1} = \left\{\begin{array}{l}
\frac{1}{12\pi}\frac{1}{\eps^2} \quad \mbox{spin 0}, \\
\frac{s}{24\pi}\frac{1}{\eps^2} \quad \mbox{spin 1/2}, \\
\frac{d-8}{12\pi}\frac{1}{\eps^2} \quad \mbox{spin 1},
\end{array}\right.
\eea
where $s$ is the dimension of fermion representation,
\footnote{For instance, $s=2$ for a Weyl spinor in 4 dimensions.}
and $\eps$ is a parameter of regularization.
For a field theory in which no fields of higher spin are present
(gravity is induced),
we need
\be
N_0+\frac{s}{2}N_{1/2}+(d-8)N_1 = 0,
\ee
where $N_0$, $N_{1/2}$ and $N_1$ are the number
of fields of spin $0$, $1/2$ and $1$, respectively.
Obviously this is impossible if $d\geq 8$,
in which case fields of higher spins are needed.

String theory is classified as the possibility (\ref{2}).
In Witten's bosonic open string field theory \cite{Witten},
all fields have the canonical kinetic term,
and they appear in the cubic interaction terms in the form of
\be
\tilde{\phi} = e^{a^2\del^2}\phi,
\ee
where $a^2 = \ln(3\sqrt{3}/4)\a'$.
Similarly, in the bosonic closed string field theory
we have $a^2 =\frac{1}{2}\ln(3\sqrt{3}/4)\a'$ \cite{KS2}.
We can rewrite the action in terms of $\tilde{\phi}$,
and then the interactions are Yukawa couplings.
The kinetic term now gets an exponential factor of derivatives.
The quadratic part of the Lagrangian is $\phi{\cal D}\phi$ with
${\cal D}$ of the form
\be
{\cal D} = (k^2+m^2)e^{2 a^2 k^2}.
\ee
This give a propagator which exponentially decays to zero at large $k^2$,
and is good for making Feynman loop diagrams finite.
But its quantum contribution to Newton constant diverges.
The conclusion is again that there is a conspiracy among the different species
in the particle spectrum of string theory,
a picture consistent with the wide spread belief that
there is big symmetry yet to be understood \cite{GM}.

String theory also serves as an example in which our assumption of
a zeroth order approximation is valid.
Although string theory has a universal coupling constant
that determines all couplings,
one can turn on the background and turn off the string coupling
at the same time such that in a double scaling limit
the coupling to the background geometry is finite.
A well known example is the Anti de Sitter space with RR flux.
The radius of the Anti de Sitter space is determined by
the flux $N$ and the string coupling constant $g_s$.
One can take the limit $g_s\rightarrow 0$
and $N\rightarrow \infty$ simultaneous such that
the radius is fixed and large.

\section*{Acknowledgment}

The author thanks Chong-Sun Chu, Herbert Fried, Jong-Ping Hsu,
Kyungsik Kang, Hsien-chung Kao, Albion Lawrence, Feng-Li Lin,
Horatiu Nastase, Sanjaye Ramgoolam, Andrew Strominger
and Nicolaos Toumbas for helpful discussions.
This work is supported in part by
the National Science Council,
the National Center for Theoretical Sciences,
the CosPA project of the Ministry of Education,
Taiwan, R.O.C.,
and the Center for Theoretical Physics
at National Taiwan University.


\vskip .8cm
\baselineskip 22pt
\begin{thebibliography}{99}
\itemsep 0pt

\bibitem{Sakharov}
A.~D.~Sakharov,
And The Theory Of  Gravitation,''
Sov.\ Phys.\ Dokl.\  {\bf 12}, 1040 (1968)
[Dokl.\ Akad.\ Nauk Ser.\ Fiz.\  {\bf 177}, 70
(1967\ SOPUA,34,394.1991\ GRGVA,32,365-367.2000)].

\bibitem{SU}
L.~Susskind and J.~Uglum,
Phys.\ Rev.\ D {\bf 50}, 2700 (1994)
[arXiv:hep-th/9401070].

\bibitem{Unruh}
W.~G.~Unruh,
Phys.\ Rev.\ D {\bf 51}, 2827 (1995).

\bibitem{CJ}
S.~Corley and T.~Jacobson,
Phys.\ Rev.\ D {\bf 54}, 1568 (1996)
[arXiv:hep-th/9601073].

\bibitem{CCL}
D.~Chang, C.~S.~Chu and F.~L.~Lin,
arXiv:hep-th/0306055.

\bibitem{HB}
N.~Hambli and C.~P.~Burgess,
Phys.\ Rev.\ D {\bf 53}, 5717 (1996)
[arXiv:hep-th/9510159].

\bibitem{LW}
F.~Larsen and F.~Wilczek,
Nucl.\ Phys.\ B {\bf 458}, 249 (1996)
[arXiv:hep-th/9506066].

\bibitem{YF}
C. N. Yang, D. Feldman,
Phys. Rev. {\bf 79}, 972 (1950).

\bibitem{CHY1}
T.~C.~Cheng, P.~M.~Ho and M.~C.~Yeh,
Nucl.\ Phys.\ B {\bf 625}, 151 (2002)
[arXiv:hep-th/0111160].

\bibitem{CHY2}
T.~C.~Cheng, P.~M.~Ho and M.~C.~Yeh,
Phys.\ Rev.\ D {\bf 66}, 085015 (2002)
[arXiv:hep-th/0206077].


\bibitem{GHR}
J.~A.~Gu, P.~M.~Ho and S.~Ramgoolam,
arXiv:hep-th/0101058.

\bibitem{BH}
R.~Brandenberger and P.~M.~Ho,
Phys.\ Rev.\ D {\bf 66}, 023517 (2002)
[AAPPS Bull.\  {\bf 12N1}, 10 (2002)]
[arXiv:hep-th/0203119].

\bibitem{HL}
Q.~G.~Huang and M.~Li,
JHEP {\bf 0306}, 014 (2003)
[arXiv:hep-th/0304203].
Q.~G.~Huang and M.~Li,
arXiv:astro-ph/0308458.

\bibitem{TMB}
S.~Tsujikawa, R.~Maartens and R.~Brandenberger,
arXiv:astro-ph/0308169.


\bibitem{Kabat}
D.~Kabat,
Nucl.\ Phys.\ B {\bf 453}, 281 (1995)
[arXiv:hep-th/9503016].

\bibitem{Witten}
E.~Witten,
Nucl.\ Phys.\ B {\bf 268}, 253 (1986).

\bibitem{KS2}
V.~A.~Kostelecky and S.~Samuel,
Phys.\ Rev.\ D {\bf 42}, 1289 (1990).


\bibitem{GM}
D.~J.~Gross and P.~F.~Mende,
Phys.\ Lett.\ B {\bf 197}, 129 (1987).

\end{thebibliography}
\end{document}